\documentclass[11pt]{article}
\usepackage{scicite}
\usepackage{times}

\usepackage{graphicx}
\usepackage{epsfig}
\usepackage{amssymb}
\newenvironment{sciabstract}{%
\begin{quote} \bf}
{\end{quote}}
\newcounter{lastnote}

\title{Hidden Diversity of Vacancy Networks in\\ Prussian Blue Analogues}

\author
{Arkadiy Simonov,$^{1,2}$ Trees De Baerdemaeker,$^{1,3}$ Hanna L. B. Bostr{\"o}m,$^{1,4}$\\
Mar{\'i}a Laura R{\'i}os G{\'o}mez,$^{5,6}$ Harry J. Gray,$^{1}$ Dmitry Chernyshov,$^{7}$\\
 Alexey Bosak,$^8$ Hans-Beat B{\"u}rgi,$^{9,10}$ and Andrew L. Goodwin $^{1,\ast}$ \\
\\
\normalsize{$^{1}$Department of Chemistry, University of Oxford, Inorganic Chemistry Laboratory,}\\
\normalsize{South Parks Road, Oxford OX1 3QR, U.K.}\\
\normalsize{$^{2}$Department of Materials, ETH Zurich, CH-8093 Zurich, Switzerland}\\
\normalsize{$^{3}$Centre for Surface Chemistry and Catalysis, KU Leuven, Celestijnenlaan 200F,}\\
\normalsize{P.O. Box 2461, 3001 Leuven, Belgium}\\
\normalsize{$^{4}$Department of Chemistry, Uppsala University, PO Box 256, SE-751 05, Uppsala, Sweden}\\
\normalsize{$^{5}$Departamento de Pol{\'i}meros, Instituto de Investigaciones en Materiales,}\\
\normalsize{Universidad Nacional Aut{\'o}noma de M{\'e}xico, Ciudad de M{\'e}xico 04510, Mexico}\\
\normalsize{$^{6}$Department of Materials Science and Metallurgy, University of Cambridge,}\\
\normalsize{27 Charles Babbage Road, Cambridge, CB3 0FS, U.K.}\\
\normalsize{$^{7}$Swiss-Norwegian Beam Lines at ESRF, 6 rue Jules Horowitz,}\\
\normalsize{ F-38043 Grenoble Cedex, France}\\
\normalsize{$^{8}$European Synchrotron Radiation Facility, 6 rue Jules Horowitz,}\\
\normalsize{ F-38043 Grenoble Cedex, France}\\
\normalsize{$^{9}$Department of Chemistry, University of Zurich, Winterthurerstrasse 190,}\\
\normalsize{CH-8057 Z{\"u}rich, Switzerland}\\
\normalsize{$^{10}$Department of Chemistry and Biochemistry, University of Berne, Freiestrasse 3,}\\
\normalsize{CH-3012 Bern, Switzerland}\\
\\
\normalsize{$\ast$To whom correspondence should be addressed;}\\
\normalsize{E-mail: andrew.goodwin@chem.ox.ac.uk.}
}
\date{}

\begin{document} 

\baselineskip24pt

\maketitle 

\begin{sciabstract}
Prussian blue analogues (PBAs) are a broad and important family of microporous inorganic solids, famous for their gas storage \cite{Kaye_2005,Chapman_2005,Zamora_2010,Hu_2011b,Karadas_2012}, metal-ion immobilisation \cite{Haas_1993,Vincent_2015}, proton conduction \cite{Ohkoshi_2010,Ono_2017}, and stimuli-dependent magnetic \cite{Ferlay_1995,Verdaguer_2004,Maurin_2009}, electronic \cite{Rykov_2015} and optical \cite{Bleuzen_2000} properties. The family also includes the widely-used double-metal cyanide (DMC) catalysts \cite{LeKhac_1998,Peeters_2013,GarciaOrtiz_2014} and the topical hexacyanoferrate/hexacyanomanganate (HCF/HCM) battery materials \cite{Wessells_2011,Lee_2014,Pasta_2014}. Central to the various physical properties of PBAs is the ability to transport mass reversibly, a process made possible by structural vacancies. Normally presumed random \cite{Ludi_1970,Mullica_1978,Roque_2007}, vacancy arrangements are actually crucially important because they control the connectivity of the micropore network, and hence diffusivity and adsorption profiles \cite{Moritomo_2009,Xiong_2016}. The long-standing obstacle to characterising PBA vacancy networks has always been the relative inaccessibility of single-crystal samples \cite{Grandjean_2016}. Here we report the growth of single crystals of a range of PBAs. By measuring and interpreting their X-ray diffuse scattering patterns, we identify for the first time a striking diversity of non-random vacancy arrangements that is hidden from conventional crystallographic analysis of powder samples. Moreover, we show that this unexpected phase complexity can be understood in terms of a remarkably simple microscopic model based on local rules of electroneutrality and centrosymmetry. The hidden phase boundaries that emerge demarcate vacancy-network polymorphs with profoundly different micropore characteristics. Our results establish a clear foundation for correlated defect engineering in PBAs as a means of controlling storage capacity, anisotropy, and transport efficiency.

\end{sciabstract}

The true crystal structures of PBAs---as of Prussian Blue itself---have long posed a difficult and important problem in solid-state chemistry because their ostensibly simple powder diffraction patterns [Fig.~1(a)] belie a remarkable complexity at the atomic scale  \cite{Keggin_1936,Buser_1977,Herren_1980}. The common parent structure is based on the cubic lattice and corresponds to the idealised composition M[M$^\prime$(CN)$_6$]. Atoms of type M and M$^\prime$ (usually transition-metal cations) occupy alternate lattice vertices and are octahedrally coordinated by bridging cyanide ions (CN$^-$) at the lattice edges [Fig.~1(b)]. There is a close conceptual parallel to the double perovskite structure \cite{Howard_2005}; indeed the key considerations of covalency and octahedral coordination geometry that stabilise perovskites amongst oxide ceramics \cite{Goodenough_1955} also favour this same architecture for transition-metal cyanides, which accounts for the chemical diversity of PBAs \cite{Sharpe_1976}. Charge balance requires that the formal oxidation states of the M and  M$^\prime$ cations sum to six, as in Cd$^{\textsc{ii}}$[Pd$^{\textsc{iv}}$(CN)$_6$] \cite{Buser_1974}.

Prussian Blue itself is a mixed-valence cyanide of iron in its 2+ and 3+ oxidation states \cite{PB,Chadwick_1966}, and so its composition cannot respect this oxidation-state sum rule. Instead the rule is circumvented by the inclusion of vacancies: the composition is well approximated by the formula Fe$^{\textsc{iii}}$[Fe$^{\textsc{ii}}$(CN)$_6$]$_{3/4}\square_{1/4}\cdot x$H$_2$O, where the symbol $\square$ represents a vacancy on the M$^\prime$-site \cite{Buser_1977}. Under typical synthesis conditions, these vacancies are filled with clusters of water molecules, which complete the coordination sphere of the M cation; hereafter we imply by the term `vacancy' the possible occupancy of the M$^\prime$-site with water. Each vacancy gives rise to a micropore of effective diameter $\sim$8.5\,\AA\ \cite{Krap_2010} that is greater than the distance between neighbouring M$^\prime$-sites ($a/\sqrt{2}\simeq7.2$\,\AA). Hence a pair of neighbouring vacancies, if present, connects to form a larger micropore \cite{Kaye_2005}. A random distribution of the quarter of vacant M$^\prime$-sites would imply bulk micoroporosity, since this vacancy fraction is larger than the percolation threshold for the face-centered cubic M$^\prime$ sublattice $(\simeq0.20)$ \cite{vanderMarck_1998}. % vacancy neighbour-pairs exist, but the resulting micoropores would necessarily extend to form a percolating network \cite{ref}, in turn ensuring bulk microporosity.
But Prussian Blue is not microporous: the single-crystal X-ray diffraction study reported in Ref.~\citenum{Buser_1977} showed that, at least for certain crystallisation conditions, vacancies tend to avoid one another by adopting a specific ordered arrangement [Fig.~1(b)]. A vacancy fraction of $\frac{1}{4}$ is actually the greatest that can support complete vacancy isolation in this way. 

PBAs with a nominal composition of M$^{\textsc{ii}}$[M$^{\prime\textsc{iii}}$(CN)$_6$]$_{2/3}\square_{1/3}\cdot x$H$_2$O (hereafter abbreviated to ``M[M$^\prime$]'') contain an even higher fraction of M$^\prime$-site vacancies than Prussian Blue \cite{Sharpe_1976,Kaye_2007}. Hence geometry dictates that these vacancies---whether distributed randomly or not---must form connected neighbour-pairs [Fig.~1(b)]. The existence and nature of any extended micropore network that then develops from these local connections depends on longer-range vacancy correlations. The collective micropore structure of PBAs is remarkably poorly understood, despite the many important properties of the family and the relevance of mass transport to these properties \cite{Kaye_2005,Pasta_2014}. So what \emph{is} known? Adsorption isotherm measurements have long indicated a significant variability in pore characteristics as a function of PBA composition \cite{Kaye_2007,RoqueMalherbe_2016}. Solid-state $^{113}$Cd NMR measurements have shown evidence of non-statistical vacancy distributions throughout the solid solution Cd[Fe$_x$Co$_{1-x}$] \cite{Flambard_2009}. Weak primitive superlattice reflections have sometimes been observed and sometimes not in the powder X-ray diffraction patterns of various PBAs; their presence has usually been interpreted as evidence for (partial) Prussian-blue-type vacancy order \cite{Grandjean_2016}. High-resolution transmission electron (HRTEM) microscopy has demonstrated the presence of vacancy chains in some Cu-PBAs as well as their absence in other Zn-containing samples \cite{Calderon_2012}. And in the one existing single-crystal diffraction study of a PBA (\emph{viz}.\ Mn[Mn]), structured diffuse scattering was observed and interpreted in terms of Warren--Cowley correlation parameters \cite{Franz_2004,Chernyshov_2010}. Taken together, these observations suggest that (i) micropore network models based on random vacancy distributions are unlikely to be realistic, and (ii) there must be substantial variability amongst the pore networks of different PBAs.

%For the Zn[Co] DMC catalysts, the pore structures of some---and only some---samples have been found to collapse on dehydration to give a rhombohedral form with substantially reduced catalytic activity, but the reasons for sample specificity are not well understood \cite{Roque_2007,Marquez_2017}. 

In this study, we have characterised vacancy correlations in a range of PBAs by growing single crystal samples, measuring their X-ray diffuse scattering patterns, and interpreting these patterns \emph{via} three-dimensional difference pair distribution function (3D-$\Delta$PDF) analysis and Monte Carlo (MC) simulations. We managed to grow single crystals by employing slow-diffusion techniques (see SI.1); despite focusing predominantly on the hexacyanocobaltates M[Co] we will nevertheless come to discuss our results in the broader context of PBAs in general.

For every crystal we tested, the corresponding single-crystal X-ray diffraction pattern contained weak but highly-structured diffuse scattering, which is the hallmark of strongly-correlated disorder \cite{Keen_2015,Welberry_2016}. Representative $(hk0)$ cuts of our diffuse scattering patterns are shown for a selection of PBAs in Fig.~2, where we include the only other single-crystal diffuse scattering pattern ever reported for a PBA---namely for Mn[Mn] (Refs.~\citenum{Franz_2004,Chernyshov_2010}). Drawing on the detailed analysis of Ref.~\citenum{Chernyshov_2010} we can be confident that the main diffuse scattering features we observe are elastic rather than inelastic in origin. Inverse Fourier transform of the normalised diffuse scattering function yields the 3D-$\Delta$PDF \cite{Weber_2012}, which clearly shows the scattering to arise predominantly from vacancy correlations (and associated structural relaxation) rather than any alkali cation or solvent inclusion (see SI.2--4). In powder X-ray diffraction (PXRD) measurements, orientational averaging acts to conceal this diffuse scattering within the background function and/or to cause it to resemble primitive superlattice reflections \cite{Chernyshov_2010}; it is in this sense that the vacancy correlations from which the diffuse scattering arises are ``hidden'' from conventional PXRD analysis.

Importantly, and quite unexpectedly \cite{Chernyshov_2010}, we find a remarkable diversity of diffuse scattering patterns amongst different PBAs. This is true even for crystals with the same nominal composition but grown in separate batches (the example in Fig.~2 is a pair of Mn[Co] crystals grown in different media). So our experimental data unambiguously show that the vacancies in PBAs are distributed in a highly non-random manner, and that these distributions can be fundamentally different for different PBA samples.

How might we understand this diversity, and what are its implications for mass transport in PBAs? To answer these questions we have developed a very simple vacancy interaction model that is nevertheless capable of explaining qualitatively the various diffuse scattering patterns observed experimentally. MC simulations driven by this set of interactions generate representative pore network configurations for each phase that can then be used to determine physical properties of relevance to mass transport and storage in PBAs. The model we have developed contains just two ingredients, each arising from simple crystal-chemical considerations. The first favours a uniform vacancy distribution---\emph{i.e.}\ such that for each M-site four of its six neighbouring M$^\prime$-sites are occupied and two are vacant. This contribution reflects Pauling's ``electroneutrality'' principle \cite{Pauling_1960}, and is consistent with earlier local-structure investigations based predominantly on Cd environment distributions in Cd[Fe$_x$Co$_{1-x}$] \cite{Flambard_2009}. The second ingredient acts to favour centrosymmetric vacancy distributions around the M-site, which we expect to be more or less important as a function of  M-site chemistry (\emph{e.g.}, crystal field effects and/or tendency for off-centering). Formally, we represent the MC energy of a given vacancy model by the expression
\begin{equation}\label{mc}
E=\sum_{\mathbf r \in \{\textsc{m}\}}\left[J_1\bigg(4-\sum_{\mathbf r^\prime \in\frac{1}{2} \langle100\rangle}e_{\mathbf r+\mathbf r^\prime}\bigg)^2+\frac{J_2}{2}\sum_{\mathbf r^\prime \in\frac{1}{2} \langle100\rangle}(e_{\mathbf r+\mathbf r^\prime}-e_{\mathbf r-\mathbf r^\prime})^2\right],
\end{equation}
where the sum is taken over all M-sites at positions $\mathbf r$, with the neighbouring M$^\prime$-site states $e_{\mathbf r\pm\mathbf r^\prime}=0$ (vacant) or 1 (present), and $J_1,J_2>0$ quantify the strength of the electroneutrality and centrosymmetry terms, respectively. The occupancy fraction $\frac{1}{N_{\textsc m^\prime}}\sum_{\mathbf r \in \{\textsc m^\prime\}}e_{\mathbf r}$ is fixed at $\frac{2}{3}$ (here $N_{\textsc m^\prime}$ is the number of M$^\prime$-sites in the MC model). We carried out a series of MC simulations driven by Eq.~(\ref{mc}) for a range of $J^\prime=J_1/J_2$ ratios and for various effective temperatures $T^\prime=T/J_2$ (see SI.5). Our results are shown in Fig.~3(a), represented in terms of the single-crystal X-ray diffuse scattering patterns calculated from an ensemble of 40 MC configurations generated at each point across an evenly-distributed mesh of suitable $J^\prime,\log(T^\prime)$ parameters.

The phase behaviour we observe from this simple MC model is remarkable for a number of reasons. Clearly the form of the diffuse scattering---and, as we will come to see, of the vacancy-network topology---is an extremely sensitive function of $J^\prime$ and $T^\prime$. This observation mirrors closely our experimental results: namely, that small variations in synthesis conditions or PBA composition are accompanied by substantial changes in the form of the diffuse scattering. Such sensitivity arises because the two interaction terms of electroneutrality and centrosymmetry actually operate in tension: they are mutually resolvable at a vacancy fraction of $\frac{1}{4}$ (leading to the ordered Prussian Blue vacancy arrangement shown in Fig.~1(b); \emph{cf} sample I in Ref.~\citenum{Buser_1977}), but become frustrated as additional vacancies are added to the system. Hence the two simple crystal-chemical considerations embedded in Eq.~(\ref{mc}) drive an unexpectedly complex configurational landscape for the $\frac{1}{3}$ M$^\prime$-site vacancy fraction of PBAs. This point is reminiscent of the effect of geometric frustration in relaxor ferroelectrics (\emph{e.g.}\ Pb(Mg$_{1/3}$Nb$_{2/3}$)O$_3$ \cite{Pasciak_2012}) and relaxor ferromagnets (\emph{e.g.}\ La(Sb$_{1/3}$Ni$_{2/3}$)O$_3$ \cite{Battle_2013}), for which the problem of 1:2 decoration of the face-centred cubic lattice is also central.

Importantly, the experimental diffuse scattering patterns given in Fig.~2 are well approximated by our MC simulations at different specific values of $J^\prime$ and $T^\prime$ [Fig.~3(a,b)]. The implication is that considerations of electroneutrality and centrosymmetry alone are sufficient to account for the basic form and diversity of diffuse scattering patterns observed experimentally. But what determines $J^\prime$ and $T^\prime$ for a given system? PBAs for which the M-site cation is Jahn-Teller active (Cu[Co]) correspond to smaller values of $J^\prime$, which is sensible because crystal field effects \cite{Sharpe_1976} must increase the relative importance of the centrosymmetry ($J_2$) term in the lattice energy. By contrast, crystal-field-inactive M-site cations correspond to larger $J^\prime$; the especially large value for Zn[Co] likely reflect the empirical propensity of Zn to adopt acentric  coordination geometries in its pseudobinary cyanide Zn(CN)$_2$ \cite{Zhdanov_1941,Shugam_1945}, rhombohedral PBAs \cite{Siebert_1981}, and related phases \cite{Marquez_2019} [Fig.~3(c)]. So variation of PBA composition allows some control over $J^\prime$, with M-site chemistry playing a more important role than that of the M$^\prime$-site. In contrast, the effective MC temperature $T^\prime$ will reflect precursor concentration and mechanism of crystal growth during synthesis (high $T^\prime\equiv$ rapid precipitation and/or high concentration). Indeed our different Mn[Co($^\prime$)] samples are associated with similar $J^\prime$ values but different $T^\prime$, with the lowest $T^\prime$ value obtained for the slowest-grown sample (gel diffusion). So, from a synthetic viewpoint, there is genuine scope for navigating $J^\prime,T^\prime$-space through judicious choice of PBA chemistry ($J^\prime$) and synthesis approach ($T^\prime$).

Just as the calculated diffuse scattering patterns are unexpectedly diverse for our MC configurations, so too are the corresponding vacancy-network structures. Despite their considerable disorder, these networks have meaningfully different physical characteristics that we will come to discuss in greater detail below. At the very simplest level, different configurations have vacancy-networks with very different coordination number and geometry distributions [Fig.~4; see also SI.8]. At low values of $J^\prime$, for example, the vacancy-network contains a large fraction of square-planar nodes; by contrast, at large $J^\prime$, one finds low-dimensional motifs based on 120$^\circ$ zig-zag chains dominating instead. High effective temperatures favour a greater diversity of network geometries and low temperatures stabilise uniform vacancy-networks and/or phase segregation. Collectively the various different scattering patterns and micropore geometries identify previously-unknown phase domains of distinct vacancy-network polymorphs, the boundaries between which are essentially hidden from conventional PXRD analysis [Fig.~3(b)]. 

%One specific real-space motif that recurs in different forms amongst these phases is that of ``ladder'' vacancies which can connect to form columnar, planar, or lattice-like micropores [Fig.~4(b)]. The physical relevance of this motif to real PBA samples is directly evident from the experimental 3D-$\Delta$PDF function itself [Fig.~4(c)] as it is in previous high-resolution transmission electron microscopy measurements \cite{Calderon_2012}. The structure of rhombohedral Zn[Co] DMCs \cite{Krap_2009} contains collapsed ladder vacancies [Fig.~4(d)], suggesting that varying degrees of ladder-vacancy preorganisation in cubic Zn[Co] samples may explain their varying susceptibilities to structural collapse on dehydration.

In Fig.~3(d) we show a range of physical quantities calculated from our MC configurations as a function of $J^\prime$ and $T^\prime$ (see SI.5). Some of these---\emph{e.g.}\ the MC energy gradient $\log(\frac{\Delta E}{\Delta T})$ or the degree of diffuse scattering localisation $L=\log\left[\frac{\sum I^2}{(\sum I)^2}\right]$---serve primarily to highlight phase boundaries, but others are particularly relevant to the transport properties of PBA phases. For example, tortuosity $\tau$ is a measure of the curvature of internal pore space \cite{Carman_1956,Bear_1988} and hence its variation suggests a range of both adsorption and transport profiles amongst PBAs \cite{Xiong_2016}. It is a particularly important quantity in the context of gas- or ion-storage materials because it affects the rate of mass transport (also known as conductance or volumetric flow rate \cite{vanderLinden_2016}):
\begin{equation}
C\propto\frac{\rho}{\tau^2},
\end{equation}
where $\rho$ is the number of vacancy neighbour-pairs per formula unit \cite{Epstein_1989}. We find $C$ varies by as much as a factor of two within the high-temperature disordered phase I and by yet another factor of two on progressing into lower-temperature phases. In other words, the variation in diffuse scattering we observe experimentally reflects a variation in pore network structure that accounts for a fourfold difference in transport rates. Even accessible pore volumes vary substantially: we calculate differences $>25$\% for this same family of configurations. Moreover, a number of vacancy-network polymorphs are strongly anisotropic and hence their transport properties will depend on orientation. So on the one hand this unexpected variability in micropore characteristics helps explain the irreproducibility and diversity of sorption and storage properties observed experimentally. And, on the other hand, it serves to highlight the clear opportunity for optimisation of transport and storage characteristics \emph{via} synthetic control over vacancy correlations---\emph{i.e.}\ defect engineering \cite{Fang_2015}. For example, the value of $C$ should be maximised for battery materials, and hence our results suggest targeting a combination of low $J^\prime$ and high $T^\prime$. Our analysis suggests the former can be achieved by using Cu$^{2+}$ as the M-site cation, and the latter by precipitating samples quickly from high-concentration precursors. This analysis is remarkably consistent with the independent empirical identification of polycrystalline Cu[Fe] (``CuHCF'') as a high-performance battery material \cite{Wessells_2011}.

Our results also identify a number of future challenges. In this study we have deliberately focussed on single-crystal PBA samples---indeed one of our key points is that PXRD is remarkably insensitive to the vacancy polymorphism of this family. So the task of establishing a robust link between vacancy correlations and (to take one example) ion-storage capacity in HCF battery components will require innovative approaches to measuring and interpreting diffuse scattering from microcrystalline samples. Serial femtosecond crystallography \cite{Zhang_2015} and/or electron diffraction \cite{Yun_2015} may help in this regard. Our analysis has also been intentionally simplistic: we have not needed to invoke the role of alkali metal inclusion in mediating vacancy stoichiometry or distributions, nor have we considered explicitly the role of variation in M$^\prime$ or the strain implications of cooperative JT distortions \cite{Ojwang_2016}. Yet these additional degrees of freedom must allow further chemical control over pore network characteristics. Indeed there is a very large number of potential variables one might use to engineer defects in PBAs---\emph{e.g.}\ concentration, $p$H, crystal growth rate and media, temperature, speciation, solubility, competing ions, chelation---and establishing rules that link these variables to vacancy polymorphs represents an enormous but worthwhile challenge. Our results raise new questions of the implications of vacancy-network polymorphism for magnetic order (\emph{e.g.} spin-glass formation \cite{Chernyshov_2010}), spin-state transitions, orbital order, and photophysics in these materials. Moreover, any mechanistic understanding of DMC catalysis will require characterisation of particle surface structure, which we now find may be substantially more complex (and varied) than previously anticipated.  And, stepping back, one might reasonably question whether a similar wealth of hidden polymorphism plays a role in the defect chemistry of other microporous phases, such as metal--organic frameworks \cite{Cliffe_2014,Liu_2019} and zeolites \cite{Brunklaus_2016}.

\section*{Methods}

\subsection*{Single crystal growth}

Single crystal PBA samples were grown using slow-diffusion methodologies. Typical preparations involved counterdiffusion of aqueous solutions of a potassium hexacyanometallate(III) and a divalent transition-metal nitrate, chloride, sulfate, or acetate. Crystals of Cd[Co], Mn[Co]$^\prime$, Mn[Fe], and Zn[Co] were grown in H-cells, while those of Mn[Co], Cu[Co], and Co[Co] were grown from silica gel (see SI.1). Care was taken not to dehydrate our samples.

\subsection*{Single crystal diffuse scattering}

Single crystal diffuse scattering measurements were carried out using the I19 beamline at the Diamond Light Source (U.K.) and the BM01 beamline at the European Synchrotron Radiation Facility (France). Each measurement involved a full sphere of data collection carried out in a single run. Bragg peaks were indexed and integrated using the package {\sc xds} \cite{kabsch2010xds}. Reciprocal space reconstruction and averaging was performed using the software {\sc meerkat} \cite{meerkat} (see SI.2).

\subsection*{3D-$\Delta$PDF analysis}

Diffuse scattering was analysed using the 3D-$\Delta$PDF method \cite{weber2012three,simonov2014experimental}. The experimental diffuse scattering was reconstructed as stated above. Then the background air scattering was subtracted by using an empty instrument run and manually selecting the optimal scale coefficient. The resulting diffuse scattering was averaged in the $m\bar3m$ Laue group using outlier rejection as described by Blessing \cite{blessing1997outlier}. Bragg peaks were removed using the ``punch and fill'' procedure: spheres of intensity around the Bragg peaks were removed to ensure omission of thermal diffuse scattering contributions from subsequent analysis. The resulting holes were filled with the median intensity from a small surrounding region of reciprocal space. Finally, the 3D-$\Delta$PDF map was calculated using fast Fourier transform. Quantitative 3D-$\Delta$PDF refinement was carried out using the program {\sc yell} \cite{simonov2014yell} (see SI.4).

\subsection*{Monte Carlo simulations and analysis}

Monte Carlo (MC) simulations were carried out using a parallel tempering approach \cite{Earl_2005} implemented within custom-written code. For each $J^\prime$ value, an ensemble of 129 configurations was generated and MC simulations carried out at a suitable distribution of temperatures $T^\prime$ (evenly spread in $\log T^\prime$). Replica exchange steps were implemented following regular intervals of successful MC steps. Configurations were equilibrated for a fixed number of epochs, and 40 configurations for diffuse scattering calculations were selected from a further production run. Diffuse scattering patterns were calculated from this ensemble of configurations with $m\bar3m$ symmetry applied. Convergence was almost universal except for a small family of polymorph II configurations at the very lowest sampled temperatures. Surface area and accessible pore volume calculations were calculated using the {\sc zeo++} code \cite{Willems_2012} for small configurations, and a related custom-written code for larger configurations (see SI.5).

%The ASA and AV were calculated by counting the number of vacancies belonging to the percolating clusters. The values of ASA and AV per cluster vere determined using the program  [Trees should know]. 

\cleardoublepage

\clearpage

\section*{Acknowledgements}
A.S. and A.L.G gratefully acknowledge financial support from the Leverhulme Trust U.K. (Grant No. RPG-2015-292), and T.D.B. acknowledges F.W.O.--Vlaanderen (Research Foundation Flanders) for a Postdoctoral Fellowship. A.S. thanks the Swiss National Science Foundation for an Ambizione Fellowship. M.L.R.G. thanks the Consejo Nacional de Ciencia y Tecnología (Mexico) for a Scholarship. A.L.G. thanks the European Research Council for funding (Grant Nos.~279705 and 788144), P.~D.~Battle (Oxford) and A. R. Overy (Oxford) for valuable discussions, and N. P. Funnell (ISIS), J. A. Hill (Courtauld) and C. S. Coates (Oxford) for assistance with single-crystal growth.
\clearpage

\noindent {\bf Fig.\ 1.} \\
{\bf Structure of Prussian Blue and its analogues.} (a) X-ray powder diffraction pattern of Mn[Co(CN)$_6$]$_{2/3}\cdot x$H$_2$O ($\equiv$ Mn[Co]), which is typical of that of most PBAs and even of PB itself. (b) The parent structure type (left) consists of interpenetrating f.c.c. arrays of M and M$^\prime$ cations (pink and blue spheres, respectively; cf the NaCl and double perovskite structure types), bridged by cyanide ions (black rods). In Prussian Blue (centre), one quarter of the M$^\prime$ sites are vacant, creating isolated micropores (green spheres). The M-cation coordination sites directed at these pores are typically occupied by bound H$_2$O, and additional water molecules occupy the centres of the micropores. In PBAs with composition M[M$^\prime$(CN)$_6$]$_{2/3}\cdot x$H$_2$O (right; abbreviated to M[M$^\prime$]), as many as $\frac{1}{3}$ of M$^\prime$ sites are vacant. There are now too many vacancies for each micropore to remain isolated; instead neighbouring pores must connect (dark green collars) to give an extended micropore network. The characteristics of this network depends on the degree and nature of vacancy correlations.
\clearpage

\noindent {\bf Fig.\ 2.} \\
{\bf Experimental structured diffuse scattering from Prussian Blue analogues.} The experimental X-ray diffraction patterns of single-crystal PBA samples contain highly structured diffuse scattering, the form of which is strongly sample- and chemistry-dependent. Reconstructed $(hk0)$ scattering planes are shown here for eight PBA samples ($-6<|h|,|k|<6$). We include data for two pairs of samples with the same nominal composition Mn[Co(CN)$_6$]$_{2/3}\cdot x$H$_2$O but prepared using two different media: the Mn[Co]$^\prime$ crystals were grown using H-cells and those of Mn[Co] were grown using silica gel diffusion. The data for Mn[Mn]$^\prime$ are those reported in Ref.~\citenum{Chernyshov_2010}. At the bottom-right corner of each panel we show the diffuse scattering pattern averaged over all squares with $\delta h,\delta k=2$ in the $(hk0)$ scattering plane. Intensities near the Bragg positions with even $h,k$ in the corners of the squares were removed (see SI.4). Note the fundamental difference in information content of these single-crystal data relative to PXRD traces of the same materials [\emph{cf}.\ Fig.~1(a)].

\clearpage

\noindent {\bf Fig.\ 3.} \\
{\bf Monte Carlo simulation results and vacancy-network phase diagram.} (a) Map of diffuse scattering calculated from MC configurations performed at a range of $J^\prime,T^\prime$ values. Superimposed on the map are the experimental plane-averaged diffuse scattering patterns of Fig.~2, positioned according to the best match in form of the scattering pattern (see SI.6,7). (b) Distribution of experimental PBA samples in $J^\prime,T^\prime$ space (left) and partitioning of the diffuse scattering map into regions of distinct vacancy polymorphs (I--VI). The vertical red line represents a hidden morphotropic phase boundary \cite{Jaffe_1954}. (c) Centrosymmetric and acentric M-site coordination geometries favoured at low and high values of $J^\prime$, respectively (green spheres indicate M$^\prime$ vacancies). The acentric coordination geometry allows migration of the M-site cation towards a pseudo-tetrahedral site (pink arrow). (d) Thermodynamic and micropore network characteristics of the $J^\prime,T^\prime$-dependent MC simulations: (clockwise from top-left) normalised MC energy per formula unit $E^\prime$; MC energy gradient $\log(\frac{\Delta E}{\Delta T})$; anisotropy $\sigma=\sum(I-\hat I)^2$, where $I$ and $\hat I$ are the diffuse scattering intensities respectively before and after inclusion of $m\bar3m$ Laue symmetry; diffuse scattering localisation $L=\frac{\sum I^2}{(\sum I)^2}$ which acts as a measure of degree of local vacancy order; fraction of surface-accessible vacancies $x_{\rm{acc}}$; conductance $C$; number of vacancy-neighbour pairs per formula unit $\rho$; and tortuosity $\tau$. The phase space represented by each panel corresponds to that shown in (a). Phase boundaries are shown as guides to the eye.

\clearpage

\noindent {\bf Fig.\ 4.} \\
{\bf Statistical properties of micropore network structures.} The vacancy--vacancy coordination number (outer circle of each pie chart) and coordination geometry (inner circle) distributions for MC configurations representative of phases I--VI. The charts are arranged to reflect the corresponding position on the $J^\prime,T^\prime$ phase diagram (bottom centre; open circles). Coordination numbers 0 and 1 correspond, respectively, to isolated M$^{\prime}$ vacancies and vacancy network `dead-ends'; the corresponding population fractions are coloured in light and dark beige. For each coordination number $\geq 2$, coordination geometries related to a square-planar micropore network of vacancies on the face-centred cubic M$^\prime$ sublattice are coloured in shades of pink; those related to a tetrahedral network are coloured in shades of green; all others are collated for a given coordination number and shown in an appropriate shade of grey.  Coordination geometries are given at the top right: empty and filled circles denote occupied and vacant M$^\prime$-sites, respectively, and bold lines show connected channels.The key point is that different vacancy polymorphs have markedly different real-space characteristics, which in turn give rise to the variety of bulk properties implied by Fig.~3(d). For example, the pore structure of phase VI is dominated by connected vacancy zig-zag chains, whereas that of phase II contains equal fractions of isolated vacancies and vacancies connected to form square-grid-like networks. Note the general preference for 90$^\circ$ pore angles at low  $J^\prime$ (left-hand side) and 120$^\circ$ angles at high $J^\prime$ (right-hand side). 

% (a) Representative micropore network structures for each polymorph. Only connected vacancies within a representative MC configuration are shown, and colour (red--blue) is used to illustrate computed transport rate along a common crystal axis (left--right). Polymorph II is clearly biphasic; the left-hand component (one third of the MC simulation box) has a vacancy fraction of 50\%, with vacancies forming planar sheets, and the right-hand component (two thirds of the MC box) has the ordered Prussian Blue structure with 25\% vacancies. The micropore network of this polymorph does not percolate along the transport axis; hence colour is used to denote independent pore structures. Polymorph VI contains one-dimensional pore channels that are not cross-linked (see inset). (b) A ``ladder vacancy'' fragment as described in the text. (c) The experimental 3D-$\Delta$PDF $g(\mathbf r)$ for Co[Co], which is representative of that for all PBAs studied here. The strongest and longest-range correlations lie along the cartesian axes (shaded), as expected for the ladder vacancy motif. (d) In the collapsed (rhombohedral) polymorph of Zn[Co], distorted vacancies also adopt ladder-type arrangements (see inset) that form intersecting columns related by the threefold rotation axes of the rhombohedral cell \cite{Krap_2009}. The degree of related vacancy order in cubic Zn[Co] (\emph{i.e.}\ with higher catalytic activity) likely influences its propensity to undergo structural collapse to this rhombohedral form on dehydration.

\clearpage

\begin{center}
\includegraphics{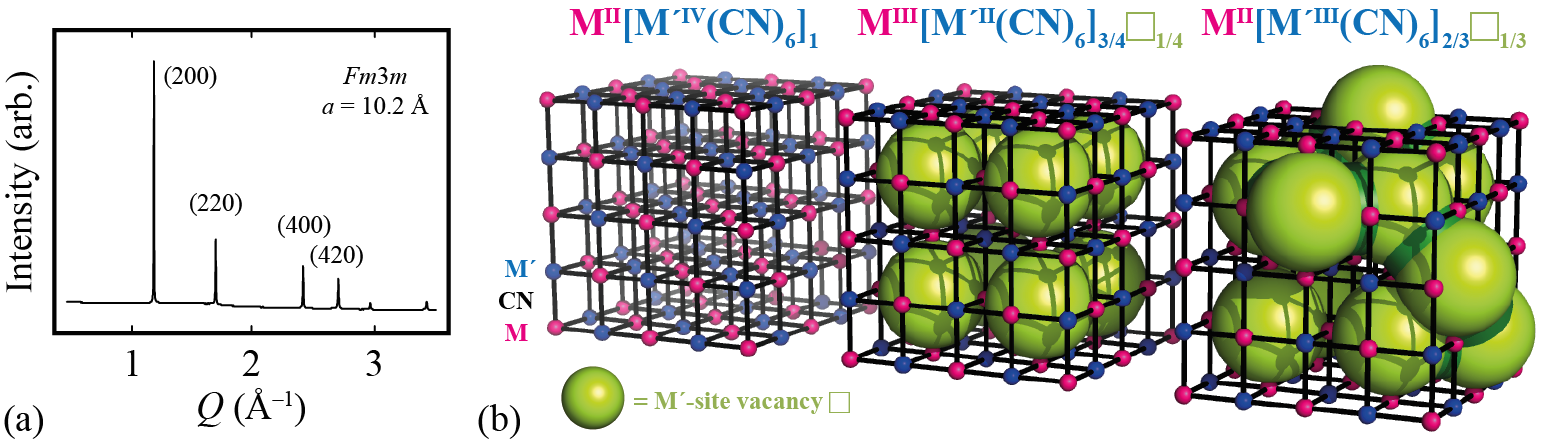}\\
FIGURE 1\\
\end{center}

\clearpage

\begin{center}
\includegraphics{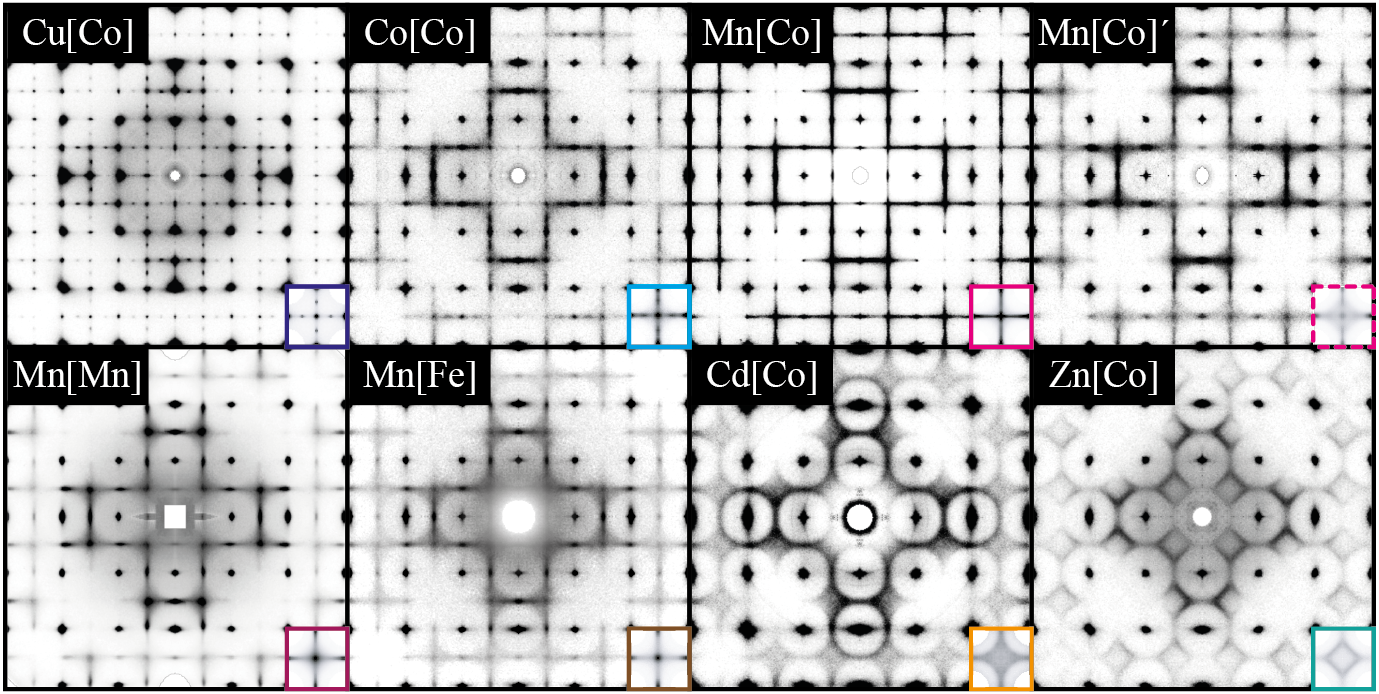}\\
FIGURE 2\\
\end{center}

\clearpage
'
\begin{center}
\includegraphics{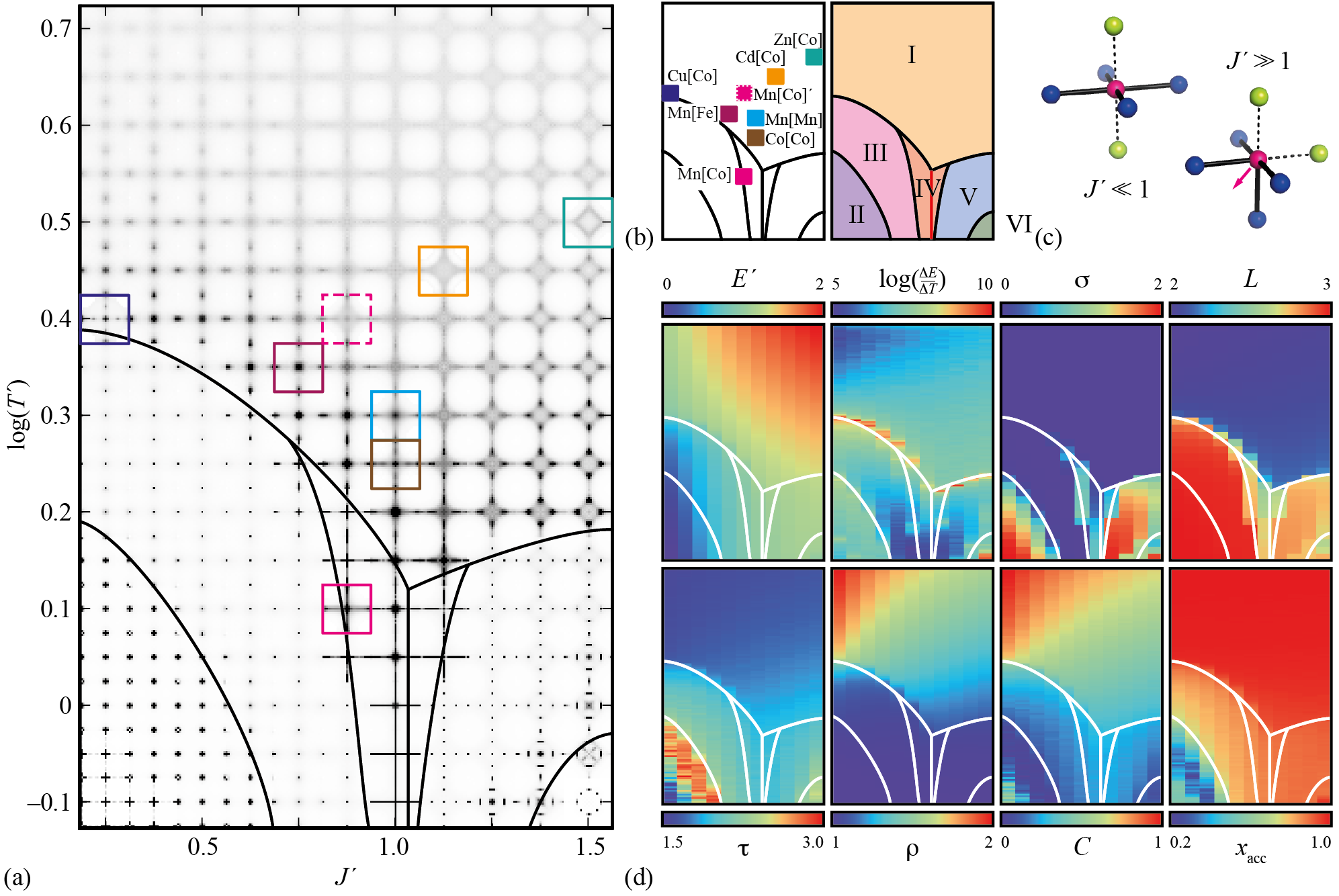}\\
FIGURE 3\\
\end{center}
\clearpage

\begin{center}
\includegraphics{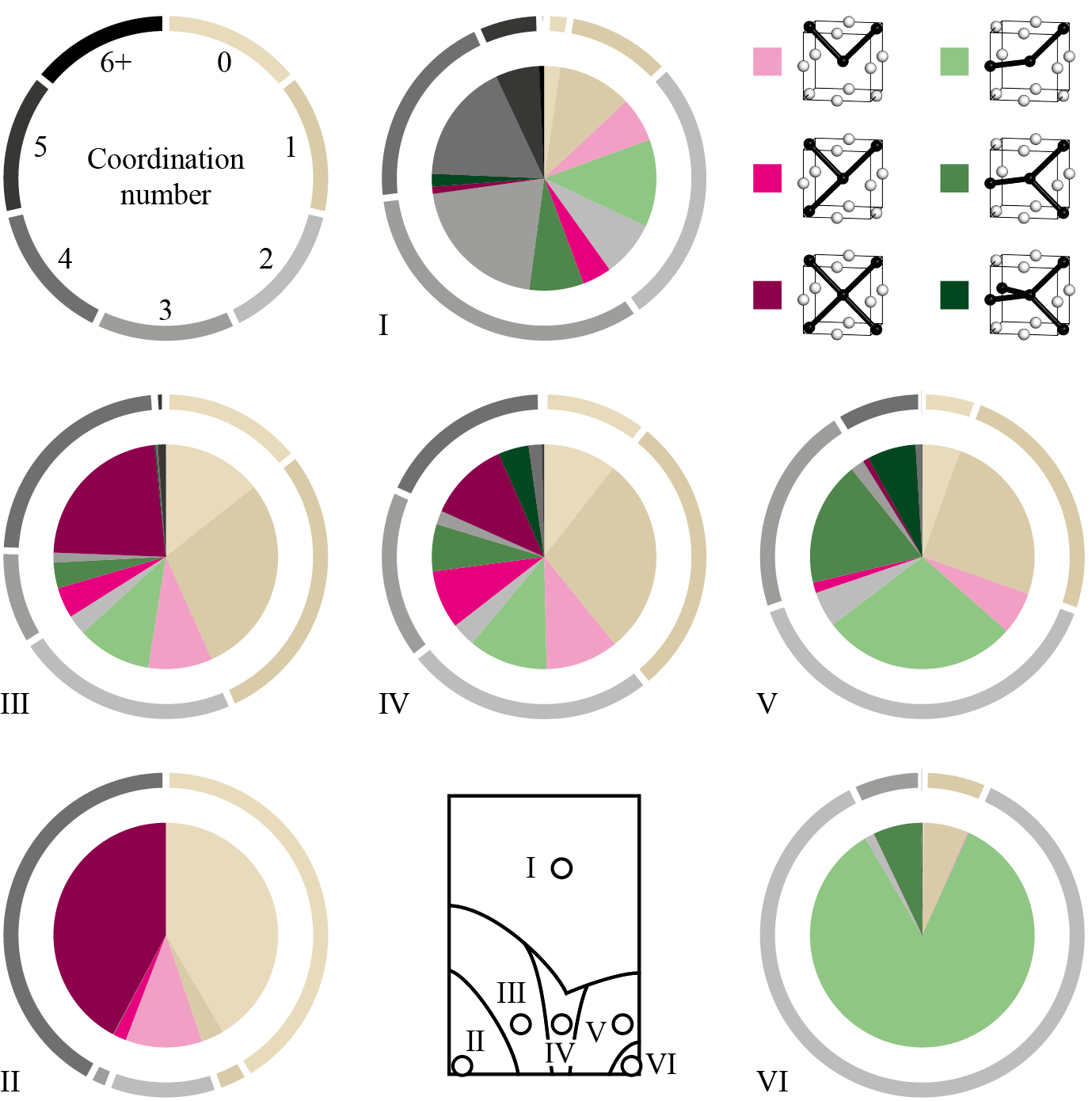}\\
FIGURE 4\\
\end{center}

\end{document}